\documentclass[a4paper,fleqn, usenatbib, useAMS]{mnras}
\usepackage{graphicx}	% Including figure files
\usepackage{amsmath}	% Advanced maths commands
\usepackage{amssymb}	% Extra maths symbols
\usepackage{multicol}        % Multi-column entries in tables
\usepackage{bm}		% Bold maths symbols, including upright Greek
\usepackage{pdflscape}	% Landscape pages

\usepackage{color, xcolor}	
\usepackage[T1]{fontenc}
\usepackage{ae,aecompl}
\usepackage{newtxtext,newtxmath, array}

\pdfoutput=1

\def\beq{\begin{equation}}
\def\eeq{\end{equation}}
\def\bey{\begin{eqnarray}}
\def\eey{\end{eqnarray}}

\title[High-energy neutrinos from TDEs]{Can tidal disruption events produce the IceCube neutrinos?}

\author[Dai \& Fang] {
Lixin Dai $^{1, 3, 4}$ \thanks{E-mail: cosimo@umd.edu (LD)}
\& Ke Fang $^{2, 3, 4}$ \thanks{E-mail: kefang@umd.edu (KF)}
\\
$^{1}$ Department of Physics, University of Maryland, College Park, MD, 20742\\
$^{2}$ Department of Astronomy, University of Maryland, College Park, MD, 20742\\
$^{3}$Joint Space-Science Institute, College Park, MD, 20742 \\
$^{4}$The authors contributed equally to this work
}
\date{}

\begin{document}
\label{firstpage}
\pagerange{\pageref{firstpage}--\pageref{lastpage}}
\maketitle

\begin{abstract}
Powerful jets and outflows  generated in tidal disruption events (TDEs) around supermassive black holes have been suggested as possible sites producing high-energy neutrinos, but it is unclear whether such an environment can provide the bulk of the neutrinos detected by the IceCube Observatory.  In this work, by considering realistic limits on the non-thermal emission power  of a TDE  jet and the  birth rate of the TDEs with jets pointing towards us, we show that it is hard to use the jetted TDE population to explain the large flux and isotropic arrival directions of  the observed  TeV--PeV neutrinos.  Therefore, TDEs cannot be the dominant sources, unless those without aligned jets can produce wide-angle emission of high-energy neutrinos. Supposing that is the case, we list a few recent jetted and non-jetted TDEs that have the best chance to  be detected by IceCube, based on their energetics, distances, and directions. A spatial and temporal association of these predicted events  with the IceCube data  should provide a decisive test on TDEs as origin of the IceCube neutrinos.  
\end{abstract}
\maketitle


\section{Introduction}
\label{sec:introduction}

The IceCube Observatory has recently reported the first high-energy cosmic neutrinos with more than $5\sigma$ significance \citep{Aartsen:2013jdh}. The all-flavor diffuse neutrino flux is analyzed to be $\Phi_\nu=7\times10^{-18}\,(E_\nu/100\,\rm TeV)^{-2.49}\,\rm GeV^{-1}cm^{-2}sr^{-1}s^{-1}$ for the energy range $27\,\rm TeV < E_\nu < 2\,\rm PeV$ under a simple power law hypothesis \citep{2015PhRvD..91b2001A}. No significant small-scale anisotropy has been found in the  data \citep{Aartsen:2014ivk} and no point-like or extended sources have been reported in all-sky searches \citep{2017ApJ...835..151A}. In addition,  no excess  has been found in searching for  neutrinos from gamma-ray bursts \citep{2016ApJ...824..115A} and Fermi blazars  \citep{Aartsen:2016lir}. 
The origin of the IceCube neutrinos remains a mystery \citep{MuraseReview}.

Among other sources, tidal disruption events (TDEs) have been suggested as possible origins of high-energy neutrinos \citep{Wang11, Pfeffer15, Wang16}. A TDE happens when a star approaches a supermassive black hole (SMBH) so close that the tidal force from the black hole exceeds the stellar self-gravity and tears apart the star \citep[e.g.,][]{Hills75, Lacy82, Carter83, Rees88}. A few dozens of TDE candidates have been found based on the characteristic optical, UV and/or X-ray emission produced as stellar debris loses orbital energy and is eventually accreted onto the black hole \citep[for reviews of observational status of TDEs, see][]{KomossaReview, Auchettl16}. While a TDE is a transient event typically lasting about one year, the rate that stellar material is supplied back to the vicinity of the black hole can be over 100 times the Eddington accretion rate for black hole of mass around $10^6~M_\odot$ \citep{Evans89, Guillochon14}.  Stellar material can circularize due to stream-stream collisions caused by the general relativistic apsidal precession of the debris orbit \citep{Shiokawa15, Bonnerot16, Hayasaki16} \citep[although Lense-Thirring precession around a Kerr black hole can delay this process, see] []{Dai13, Guillochon15}, as well as due to nozzle shocks in the debris stream during pericenter passages \citep{Kochanek94, Guillochon13}. The efficiency of circularization through stream-stream collisions is high when the black hole is relatively massive or when the stellar orbital impact parameter is large \citep{Dai15}, and in such cases the accretion rate can also be larger than the Eddington accretion rate (for observational evidence, see e.g., \citealp{Kara16, Lin17}). This implies that the channels and environment for neutrino production in TDEs could be different from those in active galactic nuclei (AGN), as the latter are usually accreting at sub-Eddington rates at low redshift.

To date three TDEs have been observed with relativistic jets pointing towards us: \textit{Swift} J1644+57 (\textit{Sw} 1644 hereafter) \citep{Bloom11, Burrows11, Levan11, Zauderer11, Zauderer13, Saxton12, Levan16}, \textit{Swift} J2058+05 \citep{Cenko12, Pasham15}, and \textit{Swift} J1112-82 \citep{Brown15}. The jet in \textit{Sw} 1644 has a Lorentz factor of $2-10$ \citep{Bloom11, Zauderer11, Berger12, Metzger12}.  \cite{Farrar14} and \cite{Pfeffer15} showed that such TDE jets can accelerate protons and nuclei to $10^{20}\,\rm eV$ and could be relevant to the ultrahigh-energy cosmic ray hotspots \citep{Abbasi:2014lda}. The environment of the TDE jet is expected to be filled with baryons of the debris outflows as well as photons, including non-thermal photons from inverse-Compton emission of the electrons accelerated by the jet, and thermal photons from the accretion disk and outflows. The interaction between cosmic rays and the photons from the jet environment can produce secondary particles including high-energy neutrinos and $\gamma$-rays.  This has been investigated in \cite{Wang11, Wang16}. In particular,  \cite{Wang16} showed that abundant non-thermal and thermal photons from the ambience can serve as compelling cosmic ray targets with a high pion production efficiency. 

A key question is whether neutrinos from jetted TDEs are sufficient to meet the IceCube observed flux. In this work, we examine the upper limit on high-energy neutrino flux that could be produced by jetted TDEs, based on realistic TDE jet power and the observed rate of jetted TDEs. We find that if the neutrino emission is limited to the jet region, the total flux from the jetted TDE population cannot meet the IceCube measurement.  Therefore, TDEs cannot be the dominant contributor to the IceCube neutrinos unless a significant fraction of them, with or without the presence of a jet like that of \textit{Sw} 1644, can emit or redistribute neutrinos into a wide angle. To test this possibility,  we calculate the expected number of neutrino events  produced by a few nearby and powerful TDEs. We show that if some of these TDEs, which do not possess jets pointing towards us, can emit neutrinos towards Earth,  their events can stand out of  the atmospheric neutrino background and be observed by IceCube. A confirmation or absence of a temporal and spatial association of these predicted events to the IceCube data will therefore strongly support or rule out the TDE proposal. 

Our work is organized as follows. In Section~\ref{sec:energetics} we discuss how much neutrino flux can be produced in TDEs from a theoretical perspective. In Section~\ref{sec:obs} we present  properties of the observed jetted TDEs, including their birth rate and  luminosity. In Section~\ref{sec:jet} we examine the energy bucket of the  jetted TDE population in producing high-energy neutrinos in light of these observational facts. In Section~\ref{sec:outsideJet} we investigate the possibility that cosmic rays and neutrinos can be produced by TDEs that do not have jets pointing towards us. In Section~\ref{sec:numbercheck} we calculate the number of neutrinos that IceCube should have detected from a few nearby, luminous TDEs, to test the proposal that TDEs are the dominant sources of high-energy neutrinos. We summarize our findings in Section~\ref{sec:discussion}. 

\section{TDE energetics for neutrino production}\label{sec:energetics}

\subsection{Constraints from TDE observations}\label{sec:obs}
The total energy bucket of a TDE is the rest-mass energy of the star.  For a star with mass $M_{\star} \sim 1 M_{\odot}$, this total available energy is 
\bey
E_{\rm tot} \simeq 10^{54}  \ {\rm erg}.
\eey
The rate of TDEs is observed to be ${\cal R}\sim10^{-5}$ per galaxy per year \citep{Donley02, VV14, Khabibullin14}, which corresponds to ${\cal R}\sim 10^{-7}\,\rm Mpc^{-3}yr^{-1}$. The rate is predicted to be higher theoretically,   ${\cal R} =10^{-4}-10^{-5}$ per galaxy per year \citep{Magorrian99, Wang04, Stone16}.  The rate of the occurrence of TDEs with pointing relativistic jets is confined to be ${\cal R}_{\rm jet}\sim 3\times10^{-11}\,\rm Mpc^{-3}\,yr^{-1}$  \citep{Farrar14, Brown15, Sun15, Levan16}, though some uncertainty exists due to the limited number of such events.

Among the three observed jetted TDEs, \textit{Sw} 1644 is the best measured in multiple wavebands.  The mass of its host SMBH has an upper limit of $M_{\rm BH} = 8 \times 10^6 M_\odot$ from the variability timescale ($\sim 100$s) of the X-ray light curve \citep{Bloom11, Burrows11}, and is further confirmed to be several million $M_\odot$ from the reverberation analysis of the X-ray light curve \citep{Kara16}. This corresponds to a Schwarzschild radius $R_{\rm s} \sim 2.4 \times 10^{12} \left( {M_{\rm BH}}/  {8\times 10^6 M_\odot} \right) \rm{cm}$, and an Eddington luminosity of:

\bey
L_{\rm Edd} \sim 10^{45} \left( \frac{M_{\rm BH}}  {8\times 10^6 M_\odot} \right) \rm{erg \ s^{-1}}.
\eey

The averaged X-ray luminosity for the first few weeks, if the emission is isotropic, is $L_X^{\rm iso}\sim 10^{47} \rm {erg \ s^{-1}}$.  Using the Eddington luminosity as an upper limit on the jet luminosity, $L_{\rm jet}\sim 10^{45}  \rm{erg \ s^{-1}}$, the opening angle of the relativistic jet is estimated to be  $\theta_{\rm j}\sim \left(L_{\rm jet}/L_X^{\rm iso}\right)^{1/2}\sim 0.1$. The Lorentz factor of the jet is then $ \Gamma \sim 1/ \theta_{\rm j}\ \sim10$. Therefore, the total X-ray energy emitted in the 1-10 keV band in the first few weeks is $E^{\rm X}_{\rm jet}\sim10^{53} f_b\,\rm erg$, for a beaming factor $f_b \sim \Gamma^{-2} \lesssim 0.01$ \citep{Bloom11}.

A jet as strong as this can be magnetically driven via the Blandford-Znajek mechanism \citep{Blandford77}, as suggested by general relativistic magnetohydrodynamics simulations \citep[e.g.,][]{McKinney06}. For an order-of-magnitude calculation, the luminosity of such a jet is linked to the magnetic flux by: $L_{\rm jet} \sim R_s^2 c \times (B^2/4\pi)$, where $B$ is the strength of the magnetic field at the base of the jet and $c$ is the speed of light. 
This leads to $B\,R \sim 10^{17}\,\left( {L_{\rm jet}}/{10^{45}\,\rm erg\,s^{-1}}\right)^{1/2}\,\rm G\ cm$ 
near the base of the jet, where $R$ is the jet radius.
Analysis of the radio signal from this source further confirms $B R \sim 10^{16-17}\rm G \ cm$ at early times \citep{Duran13}. How \textit{Sw} 1644 can accumulate magnetic flux orders of magnitude greater than the total magnetic flux from a star is an unsolved issue, which is beyond the scope of this paper.

The X-ray flux of \textit{Sw} 1644 dropped by several orders of magnitude about one year after the peak. A likely reason is that as the stellar material supply dropped with time, the disk went through a state transition. As the disk switched from a super-Eddington thick disk to a sub-Eddington thin disk, the jet either shut off or became very weak \citep{Tchekhovskoy14}. No particle acceleration should be expected beyond this point.

\subsection{Neutrino production inside the jet }\label{sec:jet}

Particle acceleration in relativistic jets has been widely discussed in the literature \citep [e.g.,][]{1978MNRAS.182..147B, 1987PhR...154....1B}. A general criterion, namely,  the Hillas condition \citep{Hillas84}, requires  the size of the acceleration region to be larger than the particle's Larmor radius. Applying this criterion to \textit{Sw} 1644 and considering that the acceleration happens over a region of size of $R/\Gamma$ in the plasma rest frame (e.g., \citealp{2009JCAP...11..009L}), a particle with charge  $Z e$ (where $e$ is the electron charge) can be accelerated by the jet up to a maximum energy:
\bey\label{eqn:Emax}
E_{\rm CR} 
\sim 3\times10^{17}\,Z\,\left(\frac{B\ R}{10^{16}\,\rm G\ cm}\right)\left(\frac{\Gamma}{10}\right)^{-1}\,\rm eV.
\eey
As a neutrino produced in a photopion interaction can take about 5\% of the proton's energy \citep{1968PhRvL..21.1016S}, the cosmic ray particles accelerated by the \textit{Sw} 1644 jet are  energetic enough to produce the IceCube PeV neutrinos.

The accelerated cosmic ray particles encounter the non-thermal photons emitted by relativistic electrons inside the jet, as well as thermal photons and gas baryons from the environment that are swallowed by the jet. Most of the cosmic rays are expected to undergo photopion interaction. The interaction probability of a cosmic ray particle with X-ray photons in the jet, or the so-called pion production efficiency $f_\pi$, can be estimated as  \citep{Wang16}:  $f_\pi \sim 0.1\,\left(L_{X}^{\rm iso}/10^{47}\,\rm erg\,s^{-1}\right)\,\left(\Gamma/10\right)^{-2}\,\left(r/10^{15}\,\rm cm\right)^{-1}\,\left({\varepsilon}_{X}/\rm keV\right)^{-1}$, where $\varepsilon_{X}$ is the rest-frame energy of the background photon. $f_\pi$ could be even smaller for a less luminous TDE. Following \citet{Wang16}, this efficiency is evaluated at a radius of $r\sim 10^{15}$ cm from the black hole because the X-ray variability timescale indicates that non-thermal photons can be produced at that distance. However, we use the isotropic jet luminosity $L_{X}^{\rm iso}\sim10^{47}\,\rm erg\,s^{-1}$ averaged over the first few weeks instead of the $10^{48}\,\rm erg\,s^{-1}$ luminosity used by \citet{Wang16}, as the latter is the luminosity in the first few days before the jet is stabilized \citep{Tchekhovskoy14}.

Below we consider the energy bucket of the TDEs in producing high-energy neutrinos. In a TDE associated with a previously quiescent SMBH, the jet can only be powered by the transient accretion of stellar material. Numerical simulations suggest that a typical magnetically driven jet in a super-Eddington system can carry $\lesssim $10\% of the accretion energy \citep{McKinney14, McKinney15}. Hence the total energy carried by the jet is $E^{\rm total}_{\rm jet} \lesssim 10^{53}\,\rm erg$. 
The equipartition hypothesis suggests an equality of the energies in relativistic particles and magnetic fields \citep{2011hea..book.....L}. Cosmic rays at the highest energies should thus carry an energy significantly less than $E^{\rm total}_{\rm jet}$. (We note, however, it is unclear whether the equipartition hypothesis strictly applies to TDE jets, see for example \citealp{2013ApJ...770..146B}). 
For \textit{Sw} 1644,  non-thermal X-rays,  which are likely produced by the inverse Compton process of electrons in the jet, carry less than 1\% of the jet energy, $E^{\rm X}_{\rm jet} / E^{\rm total}_{\rm jet} \sim 1\% $.
Even if 10 times more energy can be channeled into cosmic ray protons than electrons in the jet \citep{Wang16}, the fraction of the jet energy that is carried by protons, $\eta_{\rm CR}\equiv E^{\rm CR}_{\rm jet} /E^{\rm total}_{\rm jet}$, is still $ \lesssim 10\%$.

Therefore, if neutrinos are produced inside the jet and can only come from TDEs with jets pointing towards us, the  energy injection rate in cosmic rays from these TDEs should be strictly below:
\bey\label{eqn:budget}
{\cal \dot {E}}_{\rm CR} 
&<& 3\times  10^{43}\, \left(\frac{\eta_{\rm CR}}{10\%}\right) \left(\frac{{\cal R}_{\rm jet}}{3\times 10^{-11}\,\rm Mpc^{-3}\,yr^{-1}}\right)  \\ \nonumber
&&\left(\frac{E_{\rm jet}}{10^{53}\,\rm erg}\right) \,\left(\frac{f_b}{0.01}\right)^{-1}\,\rm erg\,Mpc^{-3}\,yr^{-1}. 
\eey
The energy injection rate required by the measured neutrino flux, assuming the source distribution with redshift follows star formation rate (e.g., \citealp{2006ApJ...651..142H}),  is \citep{2013PhRvD..88l1301M, 2013arXiv1311.0287K}
\bey\label{eqn:need}
{\cal {\dot E}}_{\rm CR,\rm obs}\sim  5\times10^{44}\, \left(\frac{f_\pi}{0.1}\right)^{-1} \,\rm erg\,Mpc^{-3}\,yr^{-1}.
\eey
This number can be  a few times higher if TDEs follow a weaker evolution history.  Equation~\ref{eqn:budget} demonstrates that  if neutrinos can only be produced inside the jet and if cosmic rays carry no more than 10\% of the jet energy, TDEs are not expected to contribute more than $\sim6\%$ of the IceCube events, given the limited available energy and event rates of the jetted TDE population.

Another problem  with the jetted TDE scenario is that its extremely rare event rate is inconsistent with the absence of strong anisotropy in the arrival directions of muon neutrinos  \citep{2017ApJ...835..151A}. No point source  has been found  in the 7-year data, implying that the source rate needs to be more abundant than $\sim10^{-9}-10^{-8}\,\rm Mpc^{-3}\,yr^{-1}$\citep{2014PhRvD..90d3005A}.

\subsection{Neutrino production by other mechanisms}\label{sec:outsideJet}
The remaining possibility that allows TDEs to have a significant contribution to the IceCube neutrinos is that the TDEs without an Earth-pointing jet  can also emit neutrinos into our line of sight. In section \ref{sec:jet}  we have only considered TDEs with jets beamed towards us.  If all TDE jets have opening angles similar to that of \textit{Sw} 1644, there are $f_b^{-1} \gtrsim 100$ times more TDEs with misaligned jets. Such transverse jets have indeed been indicated in  radio and sub-mm observations \citep{VV16, Yuan16, Lei16}. If accelerated cosmic rays could  leak out of the jet and produce  neutrino emission off  the jet axis,  ${\cal {\dot E}}_{\rm CR}$ could be boosted to the required level.

Another possible scenario is that cosmic rays could get accelerated by other components rather than the jet. \citet{Tamborra14}  showed that PeV neutrinos can be produced in AGN winds. Super-Eddington accretion induced outflows \citep{Ohsuga05, Jiang14, McKinney15, Sadowski16} or line-driven winds in TDE disks \citep{Cole15} can provide a similar environment for particle acceleration by mechanisms such as shocks or magnetic reconnection. The electromagnetic luminosity of the outflows is $L_{\rm B} \sim (B^2 / 8\pi) \times v_{\rm w} \times 4 \pi R^2$, which gives $BR \sim \sqrt{L_{\rm B} / v_{\rm w}}$ with $v_{\rm w}$ being the speed of the outflows. Therefore, charged particles can in principle be accelerated to a maximum energy of $E_{\rm CR} \sim e B R v_{\rm w} /c \sim e \sqrt{L_{\rm B} v_{\rm w}} / c$. Observations show that the speed of TDE outflows can range from a few hundred km/s \citep{Miller15, Cenko16} to a fraction of the speed of light \citep{Kara16}. The electromagnetic flux in super-Eddington wide-angle outflows can be as high as the Eddington flux \citep{McKinney16}. Taking a conservative estimate of $L_{\rm B} = \eta L_{\rm Edd}$ with $\eta_B =10^{-2}$, we have $E_{\rm CR} \sim 100~\eta^{1/2}_{B, -2} {L^{1/2}_{\rm Edd, 45}} (v_{\rm w}/100~{\rm km/s})^{1/2}$ PeV. This is sufficient to produce PeV neutrinos. Collision between high-speed unbound stellar debris and molecular clouds has also been suggested as a mechanism of producing high-energy cosmic rays and secondary particles \citep{Chen16}.

Despite the uncertainties in outside-jet scenarios, we do not expect a long delay between the TDE electromagnetic flare and neutrino production, given that  the typical dynamical time $\sim 100 R_s/c\sim 10^4\,\rm s$  is much shorter than the TDE duration. Therefore,  if the TDE population dominates the neutrino sky, there is a good chance for IceCube to have observed some of the known TDEs. In the next section, we shall choose a few recent, nearby TDEs as the optimal candidates, and predict the number of neutrinos that could have been detected by IceCube from them. The predicted number of neutrinos  along with the TDE information can serve as a direct check of the possibility that TDEs are sources of the IceCube neutrinos.

\section{Expected event numbers of the observed TDEs}\label{sec:numbercheck}
\begin{table*}
	\centering
	\caption{Properties of nearby bright TDEs with (predicted) peak time after June 2007}
	\label{tab:properties}
	\begin{tabular}{lcccccccr} 
		\hline
		name & redshift & dec & RA & peak time & emission$^\dagger$  & $E_{\rm rad}^{\rm obs}$ [erg] & ref\\
		\hline
		UGC 03317  & 0.004136 & +73:43:26.30  & 05:33:37.54 & 2010.9 & X  & $>4\times10^{49~*}$ & \text{ \cite{2016A&A...586A...9H}}  \\
		\hline
		PGC 1185375  & 0.00523 & +01:07:36.70  & 15:03:50.29 & 2010.2 & X    & $>10^{50~*}$ & \text{ \cite{2016A&A...586A...9H}}  \\
		\hline
		PGC 1190358  & 0.00766 & +01:17:33.17 & 15:05:28.75  & 2009.12 & X     & $>2\times10^{50~*}$ & \text{ \cite{2016A&A...586A...9H}}  \\
		\hline
		PGC 015259  & 0.014665 & -04:45:35.60 & 04:29:21.82 & 2010.2 & X     & $>3\times10^{50~*}$ & \text{ \cite{2016A&A...586A...9H}}  \\
		\hline
		iPTF 16fnl  & 0.0163 & +32:53:37.5 & 00:29:57.04 & 2016.8 &  O, U   & $2\times 10^{49}$ & \text{ \cite{Blagorodnova17}}   \\
		\hline
		XMMSL1 J0740-85 & 0.0173 & +85:39:31.25  & 07:40:08.2 & 2014.4 &  O, U, X, R    &  $5\times10^{50~**}$ & \text{\cite{Saxton16}}  \\
		\hline
		ASASSN-15oi  & 0.02 & -30:45:20.10 & 20:39:09.18  & 2015.8 &  O, U, X  & $ 5\times10^{50}$ &  \text{ \cite{Holoien16b}}  \\
		\hline
		ASASSN-14li  & 0.0206 & +17:46:26.44 & 12:48:15.23 & 2014.11 &  O, U, X, R  &$7\times10^{50}$ &  \text{\cite{Holoien16a}}  \\
		\hline
		ASASSN-14ae  & 0.043671 & +34:05:52.23 & 11:08:40.12 & 2014.1 &  O, U   &$1.7\times10^{50}$ &  \text{\cite{Holoien14}}  \\
		\hline
		Swift J1644+57  & 0.3543 & +57:34:58.8 & 116:44:49.92 & 2011.3 & X, R  &$ 10^{53} f_b$ &  \text{\cite{Bloom11, Burrows11}}  \\
		\hline
		\end{tabular}
		
		$^\dagger$ X -- X-ray,  O -- optical, U -- UV, R -- radio; * only including the reported X-ray luminosity; **calculated based on the light cruve in \citet{Saxton16}

\end{table*}

\begin{table}
	\centering
	\caption{Expected event number of the brightest TDEs, assuming $\alpha=2$ and $E_{\nu,\rm min}=10\,\rm TeV$}
	\label{ta:Nev}
	\begin{tabular}{lccr}
		\hline
		Name & $N_{\rm ev}^{\rm avg}$& $N_{\rm ev}^{\rm bol}\times (0.1/f_\pi)$ & $N_{\rm bg, 1^\circ}$\\
		\hline
		UGC 03317  & 5.11 & 0.20 &0.02  \\
		PGC 1185375  & 5.55 & 0.55 & $3.61\times10^{-4}$ \\
		PGC 1190358  & 3.76& 0.75 & $4.40\times10^{-4}$   \\
		PGC 015259  & 0.21 &$6.3\times10^{-2}$ & $3.42\times10^{-4}$  \\
		iPTF 16fnl  & 0.55 & $1.1\times10^{-2}$  & $1.48\times10^{-2}$ \\
		XMMSL1 J0740-85 & 0.31 &0.16 & 0.02\\
		ASASSN-15oi  & $2.51\times10^{-2}$ &  $1.25\times10^{-2}$& $6.96\times10^{-5}$  \\
		ASASSN-14li &  0.51 &0.36 & 0.01 \\
		ASASSN-14ae  &  $7.41\times10^{-2}$ &  $1.26\times10^{-2}$ & $1.53\times10^{-2}$  \\
		Swift J1644+57 & $7.19\times10^{-2}$& $7.19\times10^{-2}$  & $2.12\times10^{-2}$  \\
		\hline
	\end{tabular}
\end{table}

The total event number from a TDE can be calculated by 
\bey\label{eqn:Nev}
N_{\rm ev} =\int_{E_{\nu,\rm min}}^{E_{\nu,\rm max}}\,\frac{d\dot{N}_\nu}{dE_\nu}\,\frac{1}{\Delta\Omega_s\,D_L^2}\,{\Delta T_s} \,A_{\rm eff}(E_\nu,\delta)\,dE_\nu,
\eey
where $ {d\dot{N}_\nu}/{dE_\nu}$ is the averaged neutrino production rate over the TDE duration $\Delta T_s$,  $D_L$ denotes the luminosity distance of the source, $\Delta \Omega_s$ corresponds to the solid angle of the emission, and $A_{\rm eff}$ is the energy-dependent effective area of the detector that depends on the declination  of the event. Specifically,  we use the average effective area of the 86-string detector for $\nu_\mu$  and $\bar{\nu}_\mu$ neutrinos,  \footnote{The track-like events have sub-degree angular resolutions and    are typically  used by IceCube for point-source searches.} \citep{2014ApJ...796..109A}, but scale it to the actual size of the detector according to the event year. The lower limit of the integral is set to be $E_{\nu, \rm min}=$  10 TeV, above which the atmospheric background is relatively small.  Finally, the upper limit is determined by the maximum cosmic ray energy  calculated by equation~\ref{eqn:Emax} and taking that $E_{\rm CR}\sim 20\,E_\nu$.  

Assuming that the cosmic ray interactions lead to a production of charged pions with roughly $50\%$ probability, and that $3/4$ of the decay products of the charged pions are neutrinos, the neutrino flux can be   connected to  the cosmic ray flux by  \citep{1999PhRvD..59b3002W}:
 $E_\nu^2 {dN_\nu}/{dE_\nu} = ({3}/{8})\,f_\pi \,E_{\rm  CR}^2{dN_{\rm CR}}/{dE_{\rm CR}}$. 
The cosmic ray flux is determined by the total energy input over the TDE duration: ${\mathcal E}_{\rm CR} =  \int_{E_{\rm CR, min}}^{E_{\rm CR, max}}\,E_{\rm CR}\left({d\dot{N}_{\rm CR}}/{dE_{\rm CR}}\right)dE_{\rm CR} \,\Delta T_s$, with  $E_{\rm CR, min}\sim 10^9\,\rm eV$ being the rest mass of proton and $E_{\rm CR, max}$  defined in equation~\ref{eqn:Emax}. 

Shock acceleration generally leads to a power-law energy spectrum $dN_{\rm CR}/dE_{\rm CR}\propto E_{\rm CR}^{-\alpha}$ with the index $\alpha$  between 2 and 3  \citep{1978MNRAS.182..147B, 1987PhR...154....1B}.  
Using an energy injection rate which is constant during the TDE lifetime (a more realistic  injection rate $\propto t^{-5/3}$ barely changes the result),   and taking $\alpha=2$, equation~\ref{eqn:Nev}  can be further written as
\bey\label{eqn:Nev2}
N_{\rm ev} = \frac{3 \,f_\pi \, {\mathcal E}_{\rm CR}}{8\,\Delta \Omega_s \,D_L^2 }\ln\left(\frac{E_{\rm CR, max}}{E_{\rm CR, min}}\right)
\int_{E_{\nu,\rm min}}^{E_{\nu,\rm max}}\, {A_{\rm eff}(E_\nu,\delta)}{E^{-\alpha}_\nu}\,dE_\nu. 
\eey
In the above expression we have assumed that the earth is within the solid angle of the emission, and that  the TDE can be  observed by the IceCube Observatory for the entire event duration.  In addition, we have assumed that cosmic rays mainly interact with thermal photons in the outflows. If the background photons instead follow a spectral energy distribution, for example a band function \citep{1993ApJ...413..281B} with a break at $\epsilon_{b}\sim 1\,\rm keV$,   neutrinos below   $E_{\nu,b} =7.5 \,\epsilon_{b,\rm keV}^{-1}\,\rm TeV$ (e.g. \citealp{2014ApJ...794..126F} with $\Gamma=1$ in the outflows) could have a spectrum much softer than   the cosmic ray spectrum. Nonetheless, this  does not  change our results as long as $E_{\nu,b} < E_{\nu,\rm min}$.

As $N\propto 1/D^2\,{\cal E}_{\rm CR}$, TDEs that are the closest and most powerful have the best chances to be observed. We hence select a few nearby and energetic TDEs  that happened after the first physics run of IceCube in June 2007 \citep{2009ApJ...701L..47A}.  Their coordinates and properties are listed in  Table \ref{tab:properties}.

For the estimation of ${\cal E}_{\rm CR}$, we take two approaches. The first approach is that we shall assume every TDE is equally energetic in injecting energy into cosmic rays. Then each TDE needs to have an input energy,    
\bey\label{eqn:E_CR_avg}
\mathcal{E}_{\rm CR}^{\rm avg} = 10^{51}\,\left(\frac{\mathcal{R}}{10^{-7}\,\rm Mpc^{-3}\,yr^{-1}}\right)^{-1}\left(\frac{\dot{\cal E}_{\rm CR}}{f_\pi\,10^{44}\,\rm erg\,Mpc^{-3}yr^{-1}}\right)\,\rm erg,
\eey
such that the total injection rate of TDEs can satisfy the energy requirement of the detected neutrinos. The corresponding neutrino event is labelled as $N_{\rm ev}^{\rm avg}$ in Table \ref{ta:Nev}. Since the $f_\pi$ factor is cancelled between equations~\ref{eqn:Nev2} and \ref{eqn:E_CR_avg}, $N_{\rm ev}^{\rm avg}$ does {\it not} depend on the pion production efficiency of the source environment.

For the second approach, we consider the possibility that more luminous TDEs may produce more neutrinos. We list the constraints on the total radiation energy of TDEs in Table \ref{tab:properties}. The bolometric luminosity in an event is calculated based on the observed X-ray or optical luminosity and the assumption that the radiation has a blackbody spectrum. This leads to some uncertainties due to our limited understanding of the TDE spectra \citep{Roth16}. We then calculate the number of neutrinos detectable by IceCube assuming the energy injected into cosmic rays in these events is ten times the total energy radiated, following equation \ref{eqn:budget}.  We note that due to the lack of direct non-thermal X-ray emission from some of the objects, the bolometric luminosity serves as a limited  probe to the emission power of the source. This could be a poor assumption if the environment is very unbalanced in dissipation of non-thermal and thermal emission. This event number is labelled as $N_{\rm ev}^{\rm bol}$ in Table \ref{ta:Nev}. $N_{\rm ev}^{\rm bol}$ scales to $f_\pi$ following equation~\ref{eqn:Nev2}. We take $f_\pi\sim0.1$  for calculation, although this number in the non-jetted scenario could be significantly different.

The event numbers in Table \ref{ta:Nev} are computed assuming $\alpha=2$. The event number with a different $\alpha$ can be calculated by  $N_{\rm ev}(\alpha)=f(\alpha)\,N_{\rm ev}(\alpha=2)$, where $f(\alpha)$ is a scale factor:
\bey
f(\alpha)= \frac{   (2-\alpha)\,\left(E_{\rm CR}/E_{\nu}\right)^{2-\alpha} }{\left(E_{\rm CR,max}^{2-\alpha}-E_{\rm CR,min}^{2-\alpha}\right)} \,\frac{1}{\ln\left(E_{\rm CR,max}/E_{\rm CR, min}\right)}.
\eey  

In addition to the signal events from the TDE, we expect to see atmospheric neutrinos in the  field of view. 
The number of background events can be calculated by 
\bey
N_{\rm ev, atm} =\int_{E_{\nu,\rm min}}^{E_{\nu,\rm max}}\,\left(\frac{d\Phi_\nu}{dE_\nu}\right)_{\rm atm}\,\Delta T_s\,A_{\rm eff}(E_\nu,\delta)\,\Delta \Omega_b \,dE_\nu\,,
\eey
where $(d\Phi_\nu/dE_\nu)_{\rm atm}$ is the measured flux of the atmospheric neutrinos \citep{2015EPJC...75..116A}, and $\Delta\Omega_b$ is the solid angle of the desired observation patch along the direction of the TDE, which we take an angular scale of $1^\circ$ to be comparable to the medium angular resolution of a muon neutrino event in IceCube \citep{2014ApJ...796..109A}.
The background event number above 10 TeV is listed as $N_{\rm bg, 1^\circ}$. In order to calculate the background event number, we use 100 days instead of a typical timescale of $\sim 1$yr for the TDE duration. This is because the debris fallback decays with time as $t^{-5/3}$, so most of the material is supplied to and accreted onto the black hole in the first $\sim 100$ days in a TDE (supposing debris circularization is efficient). If energy injection traces accretion or jet power, then most of the neutrinos should also be produced in the earlier phase of a TDE.

Table~\ref{ta:Nev} suggests that a few luminous and nearby TDEs, especially UGC 03317, PGC 1185375, PGC 1190358, PGC 015259, iPTF 16fnl, XMMSL1 J0740-85 and ASASSN-14li, have a good chance to be detected. \textit{Sw} 1644, the closest of the three jetted TDEs, cannot produce neutrino flux above the background level unless $f_\pi \sim1$.  We suggest that one uses IceCube data within $\sim1$ yr around the start dates of these TDEs when searching for correlated events, due to uncertainties in the exact peak date in some events, and in order to cover possible delays between the neutrino production and onset of accretion. If TDEs are indeed the sources of high-energy neutrinos, such a targeted search with known spatial and temporal information can be more effective than an all-sky blind search \citep{2016arXiv160908027F}.

\section{Discussions}\label{sec:discussion}

We discuss in this paper the possibility that TDEs are dominant sources of the high-energy neutrinos detected by IceCube. Using upper limits on jet energy and rate of jetted TDEs obtained from observations, we find that the maximum neutrino flux that can be produced by jetted TDEs like \textit{Sw} 1644 is at least one order of magnitude below the observed flux level, and the event rate is too rare to explain the observed isotropy of muon neutrino arrival directions. Our work, however, does not rule out the possibility that jetted TDEs can contribute to a sub-fraction of the TeV--PeV neutrino events, for example, only the events at the highest energy levels. Alternatively, we consider the scenario that TDEs without jets pointing towards us can emit neutrinos in a wide angle, and list a few past TDEs that could have produced neutrinos detectable by IceCube. A search in IceCube data for temporal and spatial association with these events can either confirm the non-jetted TDE scenario, or exclude TDEs as the dominant sources of IceCube neutrinos, under the fair assumption that the observed TDEs can represent the full TDE population.

While the particle acceleration mechanism is expected to be similar inside a TDE jet and an AGN jet, a few differences can arise between the two scenarios. Firstly, the disk accretion rate in a TDE  can be super-Eddington for a significant fraction of the event duration, whereas the disk accretion rate in an AGN  is usually sub-Eddington or close to Eddington in the local universe. Therefore, the jet production mechanism might not be entirely the same in both types of events. Moreover, the photon number density in a TDE jet is likely to be higher than that in an AGN jet, resulting in more intense neutrino emission. Thirdly,  a TDE is a transient event lasting for $\sim1$ year, whereas an AGN jet can last for millions of years and therefore be a steady source of high-energy neutrinos.  This also suggests that particle acceleration sites in a TDE jet should be relatively close to the black hole.

The production of high-energy cosmic rays and neutrinos is usually accompanied by high-energy $\gamma$-ray emission. High-energy $\gamma$-rays can be produced through a leptonic channel by the electrons accelerated by the jet, and/or through a hadronic channel from the decay of neutral pions, as a side product from the same interaction that produces the neutrinos. However, so far no TDEs have been reported to be observed in high-energy $\gamma$-rays.  The reason can be that the jet medium is optically thick for high-energy $\gamma$-rays when the jet luminosity is above $\sim10^{45}\, \rm erg\, s^{-1}$ \citep{Wang16}.  As the accretion rate drops from super-Eddington to sub-Eddington in a TDE, the accretion disk changes from a geometrically thick disk to a thin disk, and the jet shuts off when the disk becomes thin. This naturally explains the disappearance of the high-energy $\gamma$-rays when the jet luminosity is lower.

\citet{Wang16} introduced a choked TDE jet model to explain the disappearance of high-energy $\gamma$-rays when the TDE jet luminosity is low. If such choked jets could exist , the rate of jetted TDEs could be increased by a large factor and the neutrino flux produced might be boosted to the observed level. This model invoked a fiducial spherically symmetric gas envelope with half solar mass around the black hole in order to be able to choke the jet and make its electromagnetic emission unobservable. However, we think that it could be hard to achieve such a set up in TDEs. There can be an envelope of outflowing gas produced during stream-stream collision or super-Eddington accretion \citep{Strubbe09, Coughlin14, Metzger16}. The gas that becomes unbound when streams collide is less than $10\%$ of the total stream mass \citep{Jiang16}, so it is probably not dense enough to choke the jet. Outflows produced in super-Eddington accretion reside between the disk inflow and the jet \citep{McKinney15}, so the relativistic, beamed jet along the direction of the black hole spin should have a low chance to collide head-on with the outflows. Therefore, we do not think we are missing a large jetted TDE population which produces enormous neutrino emission while the jet is choked by a dense gas envelope and thus remains unobserved. Lately, \citet{Generozov17} proposed that it is possible that some TDEs are born with weaker jets. When the energy of a TDE jet is at least one order of magnitude lower than that in \textit{Sw} 1644, the jet radio emission could avoid detection depending on the density of the gas in the environment. If such weak TDE jets indeed existed, they could contribute to additional neutrino production.

\section*{Acknowledgements}
{\it During the submission this paper, we were aware of independent work of \citet{Senno17} and \citet{Lunardini16} on a similar topic. We thank the authors of these papers, especially Kohta Murase, for helpful communications.}
We thank S. Bradley Cenko, Xian Chen, Erin Kara, Ruoyu Liu, Jonathan McKinney, M. Cole Miller, Nathan Roth, Nicholas Stone, Xiangyu Wang for fruitful discussions. We also thank the anonymous referee for constructive comments. LD acknowledges NASA grant TCAN NNX14AB46G and the support from the International Space Science Institute--Beijing to the team ``New Approach to Active Processes in Central Regions of Galaxies''. KF acknowledges the support of a Joint Space-Science Institute prize postdoctoral fellowship.

\bibliography{neutrinoTDE}

\bsp	% typesetting comment
\label{lastpage}
\end{document}